\documentstyle[12pt]{article}
\renewcommand{\chi}{{\cal X}}
\renewcommand{\d}{{\rm d}}
\renewcommand{\i}{{\rm i}}

\newcommand{\D}{\displaystyle}
\newcommand{\beg}{\begin{equation}}
\newcommand{\ben}{\end{equation}}

\newcommand{\fpi}{f_{\pi}}
\newcommand{\fps}{f_{\pi}^2}
\newcommand{\la}{\lambda}
\newcommand{\pau}{\partial_{\mu}}
\newcommand{\pao}{\partial^{\mu}}
\newcommand{\chicl}{\chi_{cl}}
\newcommand{\chih}{\hat{\chi}}
\newcommand{\Chi}{\left(\frac{\chi}{\chi_0}\right)}
\newcommand{\Chicl}{\left(\frac{\chicl}{\chi_0}\right)}
\newcommand{\dpri}{\prime\prime}
\newcommand{\tpri}{\prime\prime\prime}
\newcommand{\fpri}{\prime\prime\prime\prime}

\newcommand{\ovchi}{\overline{\cal X}}
\newcommand{\intc}{\int\limits_{\cal C}}
\newcommand{\rar}{\rightarrow}
\newcommand{\lra}{\longrightarrow}

\newcommand{\bigBox}{\mbox{\Large $\Box$}}
\newcommand{\Lambert}{\raisebox{-0.6ex}{\bigBox}}
\newcommand{\secspa}{\vspace*{3ex}}

\begin{document}
\hfill TPR-94-14\\

\vspace*{1.4truecm}\noindent
{\large Chiral and Gluon Condensates at Finite Temperature}

\vskip2ex\noindent
J.~Sollfrank and U.~Heinz

\vskip2ex\noindent
Institut f\"ur Theoretische Physik, Universit\"at Regensburg, 93040
Regensburg, Germany\footnote{Work supported by BMFT, DFG and a fellowship
from the Freistaat Bayern}

\vskip2ex\noindent
{\bf Abstract}\\
We investigate the thermal behaviour of gluon and chiral condensates
within an effective Lagrangian of pseudoscalar mesons
coupled to a scalar glueball. This Lagrangian mimics the
scale and chiral symmetries of QCD.

\secspa 
\noindent{1. INTRODUCTION}\\
The construction of effective models for low energy QCD is
strongly constrained by global symmetry aspects. The most important one
is the chiral symmetry which, together with its spontaneous breakdown 
at low energies, was therefore already extensively studied. 
Another symmetry of the classical QCD--Lagrangian in the
limit of vanishing quark current masses is the dilatation symmetry or
scale invariance. This symmetry exhibits an anomaly,
i.e.~it is broken on the quantum level by radiative corrections 
\cite{traceanomaly}.

An effective realisation of the QCD scale anomaly was
found in the early 80's \cite{Schechter80}, by adding to the
classical Lagrangian a scalar glueball field
$\chi$ with an interaction potential of the form 
\begin{equation}  \label{v}
V(\chi) = C\chi^4\left[\ln\left(\chi/\chi_p\right) - 1/4 \right] \; .
\end{equation}
Here $C$ and $\chi_p$ are parameters to be specified later. This
leads to $\theta^{\mu}_{\mu} = - C\chi^4 $ for the trace of the
energy--momentum tensor which is also the divergence of the dilatation
current \cite{Callan70}. This quantity with scaling
dimension 4 can then be identified with the trace anomaly in QCD,
\begin{equation}\label{i}
\langle\theta_{\mu}^{\mu\, QCD} \rangle =
\langle \frac{\beta(g)}{2g} G_{\mu \nu}^{\,a} G^{\,a \, \mu \nu} \rangle
\equiv - \langle C\chi^4 \rangle \; .
\end{equation}

A revival of this idea came with the articles of Campbell
et.~al.~\cite{Campbell90} who, somewhat controversially, suggested to
regard $\langle \chi \rangle$ as an order
parameter for a deconfining phase transition. Similar to the restoration
of spontaneous broken symmetries at high temperature and/or densities, 
deconfinement should signal itself in the effective potential (EP) of 
the $\chi$--field by a phase transition
from $\langle \chi \rangle = \chi_0 \neq 0 $ to 
$\langle \chi \rangle_{T_c} = 0$.

The breaking of the two symmetries in low--energy QCD mentioned above 
is connected with the appearance of two important
vacuum condensates, the chiral condensate
$\langle q\bar{q} \rangle$ and the gluon condensate
$\langle G_{\mu \nu}^{\,a} G^{\,a \, \mu \nu} \rangle$, respectively.
Therefore these expectation values have been considered as order
parameters for symmetry restoration.

Campbell et al.~\cite{Campbell90} conjectured a strong correlation between
the behaviour of these two condensates at rising temperature or density.
The main relation, under the assumption of factorisation, 
is \cite{Campbell90}
\begin{equation}\label{qqcam}
\langle q\bar{q} \rangle \propto \left(\frac{\langle \chi \rangle}{\chi_0}
 \right)^3  \langle {\rm Tr}( U + U^{\dag}) \rangle \; ,
\end{equation}
with $U$ denoting the chiral matrix field, defined below. This means
that a vanishing of $\langle \chi \rangle$, i.e.~of the 
gluon condensate, drives also the chiral condensate to zero.
Therefore the important question is whether the temperature scale
for vanishing of the chiral condensate is dominated by the intrinsic
chiral dynamics, which drives the disappearance of 
$\langle {\rm Tr}( U + U^{\dag} ) \rangle$,
or by the vanishing of $\langle \chi \rangle$.

The idea in \cite{Campbell90} has triggered a lot of 
(sometimes rather phenomenological) work
investigating the scaling behaviour of low energy Lagrangians at
finite temperature or density \cite{Brown91}--\cite{Dotterweich93}.
Here, we will follow the original suggestion in \cite{Campbell90}
to determine the thermal behaviour of $\langle \chi \rangle$ 
via the minimum of the effective potential, by performing a systematic
calculation of the latter. We will see that the conjectures in
\cite{Campbell90} about the character of the chiral and deconfinement
phase transitions are only partially supported by our detailed 
calculation.

We start from the nonlinear $\sigma$--model, extended by a scalar
glueball field in order to mimic the scaling properties of QCD
\cite{Campbell90}. For this effective low--energy Lagrangian we
calculate in Sec.~2 the Gaussian effective potential (GEP) for the
dilaton field $\chi$. From the GEP we can read off the temperature 
dependence of the expectation value $\langle \chi \rangle$,
leading via the identification (\ref{i}) to the
thermal behaviour of the gluon condensate. This will be done in
Sec.~3, where we also discuss the influence of thermal excitations of
the $\chi$--field on the chiral condensate.
In Sec.~4 we will discuss our results.

\secspa \noindent
{2. THE GAUSSIAN EFFECTIVE POTENTIAL}\\
We start with the same Lagrangian as in \cite{Campbell90}, but
neglecting the contribution of the $\eta^{\prime}$ meson:
\begin{equation}\label{l}
{\cal L}^{eff} = \frac{\fps}{4} \Chi^2
{\rm Tr} \left[\pau U^{\dag} \pao U \right]
+ \frac{c\fps}{2} \Chi^3 {\rm Tr} \left[{\cal M}_q(U+U^{\dag})\right]
+ \frac{1}{2}\pau\chi\pao\chi - V(\chi) \: .
\end{equation}
Here
\begin{equation} \label{u}
U=\exp\left({\rm i} \sum\limits_{a=1}^{N^2-1} 
\la^a \: \frac{\Phi^a}{\fpi} \right) \;\;\; , \;\;\;
{\cal M}_q={\rm diag} \; \left( m_u, \; m_d,\; m_s,\; \ldots \right)
\;\;\; , \;\;\; \chi_0 = \langle \chi \rangle_{T=0}, 
\end{equation}
$\la^a$ are the generators of flavour SU($N$) ($N$ is the number of quark
flavours), and $V(\chi)$ was defined in (\ref{v}).

For $\chi = {\rm const.} = \chi_0$ this Lagrangian reduces, up to an 
additative constant, to the non--linear $\sigma$--model
\cite{Sigma}. As long as the explicit chiral symmetry breaking by the
quark mass term ${\cal M}_q$ is small, terms containing higher
derivatives of $U$ can be neglected in comparison with the first
term in (\ref{l}), and (\ref{l}) is a good approximation to the chiral 
dynamics of the $N^2 -1$ Goldstone bosons of QCD at low energy.
We will thus consider the cases $N=2$ and $N=3$. 

We want to calculate the EP $V_{\rm eff}(\bar{\chi})$ with
$\bar{\chi} = \langle \chi \rangle$. We expect a breakdown of the model
in the limit $\bar{\chi} \rightarrow 0$: The expansion of the logarithmic
tree level potential $V(\chi)$ (\ref{v}) in a power series near $\chi = 0$
leads to singular coefficients for powers larger than 3. At the origin we 
will also pick up singularities in coupling constants related to the
chiral field $U$, because of the effective scaling of all $U$--terms 
in (\ref{l}) with powers of $\chi$ (see below). This means that the theory 
cannot be expanded into a Taylor series around $\bar{\chi}= 0$. 
As mentioned in \cite{Campbell90} a particle interpretation at the 
origin is thus not possible. Campbell et al.~interpreted this phenomenon
as indication that in the phase where $\bar{\chi}= 0$ the physical
degrees of freedom of the model, the Goldstone bosons $\Phi^a$ and
the glueballs $\chi$, are no longer relevant, and that this phase
corresponds to a deconfined phase where these mesons have
dissolved into their quark content.
 
The EP implements radiative corrections to the tree
level potential $V(\chi)$ arising from virtual and (at non--zero temperature)
real excitations of the degrees of freedom in the Lagrangian. Since
the latter are no longer well-defined in a state where $\ovchi=0$,
the EP cannot be perturbatively calculated near $\ovchi=0$. Thus
a direct comparison of the effective potential
(= free energy) at the origin $\bar{\chi}=0$ and at the local minimum
of $V_{\rm eff}(\bar{\chi})$ is not possible in any perturbative approach.
Therefore, it is for example impossible to decide whether the
phase transition connected with the disappearance of the scalar 
condensate $\langle \chi \rangle$ is of first or second order. This
point was apparently missed in Ref.~\cite{Campbell90}. However,
we can still calculate the shift of the local minimum of
$V_{\rm eff}(\bar{\chi})$ at $\ovchi \neq 0$ as a function of temperature 
and investigate at which temperature scale visible thermal effects
set in which could indicate the disappearance of the gluon condensate.

We also encounter another familiar problem connected with the EP
in the case of broken symmetries. The standard loop expansion
\cite{Dolan74} for the EP leads to complex contributions for values of
$\bar{\chi}$, for which the tree level potential
$V(\ovchi)$ is concave, i.e.~its second derivative is negative. This problem
has triggered a large amount of work, ranging from an interpretation of the
imaginary part as a sign for instability for the corresponding
$\ovchi$ state \cite{Weinberg87} to techniques which avoid the complex
contributions altogether \cite{Resummation,Fujimoto83}.

In \cite{Weinberg87} values of $\ovchi$ for which $V_{\rm eff}(\ovchi)$
was complex were interpreted as false vacua, and the imaginary part was 
related to the decay rate per unit volume of the false vacuum. 
One knows, however, from the definition of the effective potential that 
the exact EP is real \cite{Coleman}. Therefore we regard the
appearence of complex contributions as a sign for an unsuitable
expansion scheme. We are thus led to choose an approximation where all 
expansion coefficients are real.

We choose a variational ansatz, the Gaussian effective potential (GEP),
which was extensively studied by Stevenson et al.~\cite{Gauss}.
Most other methods, which avoid the imaginary part, are only valid
in distinct temperature regions. The interpolated loop expansion
\cite{Fujimoto83} is only tractable for low temperatures \cite{Rivers84},
while resummation techniques are valid only for high temperatures
\cite{Resummation}. With the GEP ansatz
we are able to cover the whole temperature region. A more detailed
discussion of the problem of the imaginary part and a comparison
of different methods to solve the problem is given in \cite{Sollfrank94}.

The principle of the GEP is the introduction
of an arbitrary mass parameter $\Omega$ for the dilaton field $\chi$.
The effective potential is then be minimized with respect to this
parameter, allowing only for real values of $\Omega$. 
This is in analogy to the Ritz variational principle in
quantum mechanics, where the ground state energy is the minimal energy
under variation of the wave function. We will introduce the variational
parameter $\Omega$ later within the path integral formalism
by a method developed by Okopi\'nska \cite{Okopinska87}.

We start from the finite temperature partition function $Z^{\beta}$. 
After expanding the chiral field $U$, Eq.~(\ref{u}), in terms up to fourth 
order in the Goldstone fields $\Phi$ or $\pau\Phi$, we get for the 
partition function
\beg \label{Z2}
Z^{\beta}[J^a,K] \! =
         {\cal N} \int \prod\limits_{a=1}^{N^2-1}{\cal D}[\Phi^a]\,
         {\cal D}[\chi]
        \exp \left\{ {\rm i}\int\limits_{\cal C} \d^4 x\: \left[
         {\cal L}^{\rm eff}( \Phi^a,\pau\Phi^a, \chi,\pau \chi)
        + J^a\Phi^a + K\chi \right] \right\} .
\ben
${\cal N}$ is a normalisation constant, and ${\cal L}^{\rm eff}$ is
given as
\beg \label{Leff}
\begin{array}{rcll}
\D {\cal L}^{\rm eff} & = & \multicolumn{2}{l}{
                        \D \Chi^3 \fps c \sum\limits_{q} m_q } \\[3ex]
                  &   & \multicolumn{2}{l}{
                        \D + \: \frac{1}{2} \Chi^2 \pau \Phi^a \pao \Phi^a
                        \, - \, \frac{1}{2} \Chi^3 m_a^2 \Phi^{a2} }\\[3ex]
                  &   & \D + \: {\rm Tr} \Bigg[ \frac{1}{48 f_{\pi}^2}
                        \Chi^2 \sum\limits_{abcd}^{N^2-1}
                        \la^a \la^b \la^c \la^d
                          &
                        \D  \Big( \pau\Phi^a \, \Phi^b \, \pao\Phi^c \,
                                  \Phi^d -
                                  \pau\Phi^a \, \Phi^b \, \Phi^c \,
                                  \pao\Phi^d  \\[3ex]
                 &    &    &\D  - \Phi^a \, \pau\Phi^b \, \pao\Phi^c \,
                                  \Phi^d +
                                  \Phi^a \, \pau\Phi^b \, \Phi^c \,
                                  \pau\Phi^d \Big) \Bigg]  \\[3ex]
                 &    & \multicolumn{2}{l}{
                        \D + \: {\rm Tr} \Bigg[ \frac{2 c}{48 f_{\pi}^2}
                        \Chi^3 \sum\limits_{abcd}^{N^2-1}
                        {\cal M}_q \la^a \la^b \la^c \la^d
                        \Phi^a \Phi^b \Phi^c \Phi^d \Bigg] } \\[3ex]
                 &    & \multicolumn{2}{l}{
                        \D + \: \frac{1}{2} \pau \chi \pao \chi \, - \,
                        V(\chi)\; .}
\end{array}
\ben
$J^a$ and $K$ are external sources for the Goldstone bosons and
dilaton fields, respectively. The first term is independent of $\Phi^a$. 
Therefore we add this term to the potential $V(\chi)$ and define
\beg\label{vphi}
V^{\Phi}(\chi) := V(\chi) - g_{\chi}\chi^3 
\;\;\;\;\; {\rm with} \;\;\; 
g_{\chi} := \frac{\fpi c}{\chi_0^3} \sum\limits_q m_q \; .
\ben

We will use the real--time formalism. Therefore the integration
of $x_0$ is along the real--time path $\cal C$ in the complex plane
\cite{Umezawa}. 
In order to evaluate (\ref{Z2}) we will use the sattle point approximation.
This means that we expand the dilaton field $\chi$ around
the solution of the classical equation of motion
\beg \label{cleom}
\pau\frac{\partial {\cal L}(\chi_{cl},\Phi)}{\partial(\pau\chi_{cl})}
- \frac{\partial {\cal L}(\chi_{cl},\Phi)}{\partial \chi_{cl}} = K\: .
\ben
In (\ref{cleom}) we substitute the field $\Phi$ by its expectation
value $\langle \Phi(x) \rangle = 0$. Further we define
the dilaton fluctuating field $\chih := \chi - \chicl$ and transform 
the measure for $\chi$ to
${\cal D}[\chi] = {\cal D}[\chih]$. Because of the starting
Lagrangian (\ref{l}) which contains no $\chi$-- independent terms 
the kinetic term for the meson fields scales with $(\chicl/\chi_0)^2$:
\beg\label{prophi}
{\cal L}_{\Phi {\rm kin}} =  - \: \frac{1}{2}\Phi^a \Chicl^{\!2}
                        \left[\Lambert +
                          m_a^2\left(\frac{\chicl}{\chi_0}
                        \right)\right] \Phi^a \: ,
\ben
where $m_a$ is the meson mass in the non--linear sigma model
(see eq.~(\ref{mesonmasses}) below).
In order to obtain a standard kinetic term for $\Phi^a$ we transform the 
meson fields $\Phi^a$ by
\beg
\Phi^a \longrightarrow \frac{\chi_0}{\chicl} \: \Phi^a \: .
\ben
This transformation leads to an infinite Jakobi determinant for the
integration measure of the $\Phi$  field,
\beg
{\cal D}[\Phi^a] \longrightarrow {\cal D}[\Phi^a]\times
{\rm Det}\left[\left(\frac{\chi_0}{\chicl}\right)\delta^4(x-y)\right] \: ,
\ben
which we absorb in the normalisation $\cal N$. After these manipulations
we can write the partition function as
\begin{eqnarray}
Z^{\beta}[J^a,K] &\!\!\! = & \!\!\! {\cal N} \!
        \int \prod\limits_{a=1}^{N^2-1}{\cal D}[\Phi^a]\,
        {\cal D}[\hat{\chi}] \times \nonumber\\[2ex]
&& \!\!\!\!\!\!\!\! \exp \left\{ {\rm i}\int\limits_{\cal C} \! \d^4 x \Big[
         {\cal L}_{\chicl} + {\cal L}_{\Phi {\rm kin}} +
         {\cal L}_{\Phi {\rm int}} + {\cal L}_{\chih {\rm kin}} +
         {\cal L}_{\hat{\chi} {\rm int}} + {\cal L}_{\hat{\chi}\Phi {\rm int}}+
         J^a\Phi^a + K\chicl \Big]\right\}\! ,
        \nonumber \\  \label{Z3}
\end{eqnarray}
with
\newcommand{\spacek}{3ex}
\begin{eqnarray}
{\cal L}_{\chicl}  & = & \frac{1}{2}\pau\chicl\pao\chicl \: - \:
                         V^{\Phi}(\chicl)\: ,\\[\spacek]
{\cal L}_{\Phi {\rm kin}}   & = & -\: \frac{1}{2}\Phi^a
                        \left[\Lambert + M_a^2(\chicl)\right] \Phi^a \: ,
                        \label{e1} \\[-2ex]
{\cal L}_{\Phi {\rm int}}  &=& \begin{array}{ll}
                        &\\[1ex]
                         \D {\rm Tr} \Bigg[ \frac{1}{48 f_{\pi}^2}
                        \left(\frac{\chi_0}{\chicl}\right)^2
                        \sum\limits_{abcd}^{N^2-1}
                        \la^a \la^b \la^c \la^d
                          &
                        \D  \Big( \pau\Phi^a \, \Phi^b \, \pao\Phi^c \,
                                  \Phi^d -
                                  \pau\Phi^a \, \Phi^b \, \Phi^c \,
                                  \pao\Phi^d  \\[1ex]
                         &\D    - \Phi^a \, \pau\Phi^b \, \pao\Phi^c \,
                                  \Phi^d +
                                  \Phi^a \, \pau\Phi^b \, \Phi^c \,
                                  \pau\Phi^d \Big) \Bigg]
                        \end{array} \nonumber \\[2ex]
                &&      + \: {\rm Tr} \Bigg[ \frac{2 c}{48 f_{\pi}^2}
                        \left(\frac{\chi_0}{\chicl}\right)
                        \sum\limits_{abcd}^{N^2-1}
                        {\cal M}_q \la^a \la^b \la^c \la^d
                        \Phi^a \Phi^b \Phi^c \Phi^d \Bigg] \: ,\label{e2}\\[5ex]
{\cal L}_{\chih {\rm kin}}&=&  -\: \frac{1}{2}\chih\Bigg\{ \Lambert +
                        \underbrace{ (V^{\Phi})^{\dpri}(\chicl)}_{=:
                                                        \:M_{\chi}^2(\chicl)}
                        \Bigg\} \chih \: ,\\[\spacek]
{\cal L}_{\chih {\rm int}}&=& - \: \frac{1}{6} (V^{\Phi})^{\tpri}(\chicl) 
                                                                   \chih^3
                        \: - \: \frac{1}{24} V^{\fpri}(\chicl)\chih^4
                        \:+\:{\cal O}(\chih^5) \:+ \ldots \: ,\\[\spacek]
{\cal L}_{\chih\Phi {\rm int}}&=&
                        \frac{1}{\chicl}\chih\pau\Phi^a
                        \pao\Phi^a \:-\: \frac{3}{2}\frac{1}{\chicl}
                        M_a^2(\chicl) \chih\Phi^{a\,2} \nonumber\\[2ex]
                &&      \D + \:\frac{1}{2}\frac{1}{\chicl^2}\chih^2\pau\Phi^a
                        \pao\Phi^a \:-\: 
			\frac{3}{2}\frac{M_a^2(\chicl)}{\chicl^2}
                        \chih^2\Phi^{a\,2} \: - \:
                        \frac{1}{2}\frac{M_a^2(\chicl)}{\chicl^3}
                        \chih^3\Phi^{a\,2}
                        \nonumber\\[2ex]
                &&      \begin{array}{ll}
                         \D + \:{\rm Tr} \Bigg[ \frac{1}{48 f_{\pi}^2}
                        \frac{1}{\chicl^2}\chih^2 \sum\limits_{abcd}^{N^2-1}
                        \la^a \la^b \la^c \la^d
                          &
                        \D  \Big( \pau\Phi^a \, \Phi^b \, \pao\Phi^c \,
                                  \Phi^d -
                                  \pau\Phi^a \, \Phi^b \, \Phi^c \,
                                  \pao\Phi^d  \\[1ex]
                         &\D    - \Phi^a \, \pau\Phi^b \, \pao\Phi^c \,
                                  \Phi^d +
                                  \Phi^a \, \pau\Phi^b \, \Phi^c \,
                                  \pau\Phi^d \Big) \Bigg]
                        \end{array} \nonumber \\[2ex]
                &&      + \: {\rm Tr} \Bigg[ \frac{2 c}{16 f_{\pi}^2}
                        \frac{1}{\chicl\chi_0}\chih^2
                        \sum\limits_{abcd}^{N^2-1}
                        {\cal M}_q \la^a \la^b \la^c \la^d
                        \Phi^a \Phi^b \Phi^c \Phi^d \Bigg] \nonumber \\[2ex]
                &&     + \: {\rm Tr} \Bigg[ \frac{2 c}{48 f_{\pi}^2}
                        \frac{1}{\chicl^2\chi_0}\chih^3
                        \sum\limits_{abcd}^{N^2-1}
                        {\cal M}_q \la^a \la^b \la^c \la^d
                        \Phi^a \Phi^b \Phi^c \Phi^d \Bigg] \; .\label{e3}
\end{eqnarray}
At this stage of the calculation we have for the mass of the $\chi$--field
\beg
M_{\chi}^2(\chicl) = (V^{\Phi})^{\dpri}(\chicl)\:.
\ben
The meson masses scale with $\sqrt{\chicl/\chi_0}$, and we have defined
\beg\label{skalmass}
M_a^2(\chicl) := m_a^2\frac{\chicl}{\chi_0} \: .
\ben
Further we combine the interaction terms by
\beg\label{deflint}
{\cal L}_{\rm int} = {\cal L}_{\Phi {\rm int}} +
{\cal L}_{\hat{\chi} {\rm int}} + {\cal L}_{\hat{\chi}\Phi {\rm int}}\: .
\ben

We now introduce the variational parameter $\Omega$ by the method
developed in \cite{Okopinska87}. This is done by
adding and subtracting the quadratic term $\Omega^2 \:\chih^2/2$ to the
Lagrangian (\ref{e3}). We now regard $ - \: \Omega^2 \:\chih^2/2$ as a new
mass term for the $\chih$--field and get therefore an additional contribution
to the interaction part
\beg\label{deflintom}
{\cal L}_{\rm int} \lra {\cal L}_{\rm int}^{\Omega} = {\cal L}_{\rm int} -
\frac{1}{2}(M^2_{\chi} - \Omega^2) \chih^2 \; .
\ben
This corresponds to a two point vertex, indicated by a cross in the
Feynman diagrams (see Fig.~1). We thus have to evaluate the expression
\begin{eqnarray}
Z^{\beta}[J^a,K] &\!\!\! = & \!\!\! {\cal N} \!
        \int \prod\limits_{a=1}^{N^2-1}{\cal D}[\Phi^a]\,
        {\cal D}[\hat{\chi}] \times \label{Z4}\\[2ex]
&& \!\!\!\! \exp \left\{ {\rm i}\int\limits_{\cal C} \! \d^4 x \Big[
         {\cal L}_{\chicl} + {\cal L}_{\Phi {\rm kin}} -
         \frac{1}{2} \chih\left(\Lambert + \Omega^2 \right) \chih +
         {\cal L}_{\rm int}^{\Omega} +
         J^a\Phi^a + K\chicl \Big]\right\}\! .
        \nonumber 
\end{eqnarray}
This expression (\ref{Z4}) is by construction independent of the arbitrary 
parameter $\Omega$. But in practice, in order to evaluate (\ref{Z4}), we 
have to expand the exponential containing the interaction part
${\cal L}_{\rm int}^{\Omega}$ and truncate the series at some point. 
This truncation introduces a dependence on $\Omega$. Given such an 
approximation to the EP as a function of $\Omega$, one then uses the 
principle of minimal sensitivity with respect to the arbitrary parameter 
$\Omega$ to fix $\Omega(\ovchi)$ as a function of $\ovchi$ by the
condition
\beg\label{var}
\frac{\d V_{\rm eff}^{\rm G}(\ovchi,\Omega)}{\d\Omega} = 0 \; .
\ben

In our present work we expand
$\exp \{ {\rm i} \int \d^4 x {\cal L}_{\rm int}^{\Omega}\}$
up to first order. This means that we only include vacuum bubbles
containing at most one vertex. The Feynman diagrams with one vertex
which appear in our calculation are shown in Fig.~1.

We investigate three cases. Case 1 corresponds to a pure glueball scenario, 
neglecting in (\ref{l}) all $U$--field terms. In case 2 we include a 
SU($N$) flavour representation with $N$ equal masses for the $U$--fields.
In case 3 we consider a broken SU(3) flavour representation with
$m_s \neq m_{u,d}$, denoted by SU(3)${}_{\rm b}$.
For the last case the physical meson masses in zeroth order are
\begin{eqnarray}
m_{\pi}^2 = & 2c \,m_q & = m_1^2 = m_2^2 = m_3^2 \: ,\nonumber\\
m_{K}^2  = & c (m_q + m_s) & = m_4^2 = m_5^2 = m_6^2 = m_7^2 \: ,\nonumber \\
m_{\eta}^2 = & 2c (m_q + 2m_s)/3 & = m_8^2 \: , \label{mesonmasses}
\end{eqnarray}
where $m_q = m_u = m_d$.

The result for the partition functions for the three cases,
neglecting the normalisation, is
\begin{eqnarray}
\D \left. Z^{\beta}_{\chi}[K] \right|_{J^a=0}& \!\!\!\! =\!\!\!&
                \D \exp \Bigg\{\i\intc\d^4 x \;\Bigg[
                  \frac{1}{2}\pau\chicl\pao\chicl - V(\chicl)  +
                  K\chicl
                  \nonumber\\[1ex]
              &&\begin{array}{rl}
\hspace{1.8cm}&\D  - \; I_1(\Omega^2,\,T)
                    - \frac{1}{2} (V^{\dpri}(\chicl) - \Omega^2)
                       I_0(\Omega^2,\,T) \\[2ex]
              &\D  - \; \frac{1}{8} V^{\fpri}(\chicl)
                   I_0(\Omega^2,\,T)^2 \Bigg]\Bigg\},
           \end{array} \hspace{20mm}  \label{zchi} (\theequation)
                       \nonumber \\[5ex] \refstepcounter{equation}
\D \left. Z^{\beta}_{\rm SU(N)}[K]\right|_{J^a=0}& \!\!\!\!=\!\!\! &
                \D \exp \Bigg\{\i\intc\d^4 x \:\Bigg[
                  \frac{1}{2}\pau\chicl\pao\chicl - V^{\Phi}(\chicl)  +
                  K\chicl
                  \nonumber \\[1ex]
              &&\begin{array}{rl}
\hspace{1.8cm}&\D  - \; I_1(\Omega^2,\,T)
                    - (N^2-1) I_1(M^2_{\pi}(\chicl),\,T) \\[2ex]
              &\D  - \; \frac{1}{2} [(V^{\Phi})^{\dpri}(\chicl) - \Omega^2]
                        I_0(\Omega^2,\,T)
                    - \frac{1}{8} V^{\fpri}(\chicl)
                   I_0(\Omega^2,\,T)^2 \\[2ex]
              &\D  - \; \frac{(N^2-1) m_{\pi}^2}{4N\fps}
                        \left(\frac{\chi_0}{\chicl}\right)
                   I_0(M^2_{\pi}(\chicl),\,T)^2 \\[2ex]
              &\D  - \; \frac{(N^2-1) m_{\pi}^2}{2\chicl\chi_0}
                   I_0(M^2_{\pi}(\chicl),\,T)
                   I_0(\Omega^2,\,T) \Bigg]\Bigg\}\: ,\label{zsuz}
              \quad\quad\quad\;\:  (\theequation)
           \end{array} \nonumber \\[5ex] \refstepcounter{equation}
\D \left. Z^{\beta}_{\rm SU(3)_{\rm b}}[K]\right|_{J^a=0}&\!\!\!\!=\!\!\!&
                \D \exp \Bigg\{\i\intc\d^4 x \: \Bigg[
                \D \frac{1}{2}\pau\chicl\pao\chicl - V^{\Phi}(\chicl)  +
                  K\chicl
                  \nonumber\\[2ex]
              &&\begin{array}{rcl}
\hspace{1.53cm}&\multicolumn{2}{l}{ \D  - \; I_1(\Omega^2,\,T)}\\[2ex]
              &\multicolumn{2}{l}{ \D  - \; 3 I_1(M^2_{\pi}(\chicl),\,T)
                   - 4 I_1(M^2_{K}(\chicl),\,T)
                   -   I_1(M^2_{\eta}(\chicl),\,T)} \\[2ex]
              &\multicolumn{2}{l}{ \D  - \; \frac{1}{2}
                   [(V^{\Phi})^{\dpri}(\chicl) - \Omega^2] I_0(\Omega^2,\,T)
                    - \frac{1}{8} V^{\fpri}(\chicl)
                   I_0(\Omega^2,\,T)^2} \\[2ex]
              &\D  - \; \frac{1}{24\fps} \left(\frac{\chi_0}{\chicl}\right)
                   \Big[ &
                             9m_{\pi}^2 I_0(M^2_{\pi}(\chicl),\,T)^2\\[1ex]
              &&\D      -(m_{\eta}^2 + 4cm_s) I_0(M^2_{\eta}(\chicl),\,T)^2
                        \\[1.5ex]
              &&\D             -6m_{\pi}^2 I_0(M^2_{\pi}(\chicl),\,T)
                                        I_0(M^2_{\eta}(\chicl),\,T)\\[1.5ex]
              &&\D             +16m_{K}^2 I_0(M^2_{K}(\chicl),\,T)
                                        I_0(M^2_{\eta}(\chicl),\,T)
                    \Big] \\[2ex]
              &\multicolumn{2}{l}{\D - \; \frac{1}{2\chicl\chi_0}
                   I_0(\Omega^2,\,T)
                   \Big[ 3m_{\pi}^2 I_0(M^2_{\pi}(\chicl),\,T)}\\[2ex]
              &\multicolumn{2}{l}{\D \hfill
                + \:4m_{K}^2 I_0(M^2_{K}(\chicl),\,T)
                + m_{\eta}^2 I_0(M^2_{\eta}(\chicl),\,T)
                   \Big] \Bigg]         \Bigg\}. }
           \end{array}\nonumber\\&& \label{zsud}
           \hspace{11.8cm}
           (\theequation)
           \refstepcounter{equation} \nonumber
\end{eqnarray}
The integrals $I_n$ are defined in the appendix.
From the functional $Z[K] = \exp({\rm i}W[K]) $ we get
the EP by first making a Legendre transformation
\beg \label{legen}
\Gamma[\ovchi(x)] = W[K] - \int\d^4x K(x)
\ovchi(x) \: ,
\ben
with
\beg\label{defovchi}
\ovchi(x) = \frac{\delta W[K]}{\delta K(x)} \: .
\ben
The expansion of the effective action $\Gamma[\ovchi(x)]$
in powers of derivatives
\beg\label{veff2}
\Gamma[\ovchi(x)] = \int\d^4x \left[-V_{\rm eff}(\ovchi) +
\frac{1}{2} \, Z(\ovchi) \: \pau\ovchi(x)\pao\ovchi(x)\: ... \right] \: .
\ben
contains as its first term the EP.

We apply this procedure to the partition function (\ref{zchi})--(\ref{zsud}).
The integrals $I_n$ (see appendix)
contain two parts, the infinite $T=0$ contribution and the finite
$T>0$ contribution. The infinite $T=0$ part requires a renormalisation
procedure. In a strict sense, the starting Lagrangian (\ref{l}) is not
renormalizable because of the logarithmic interaction potential
(\ref{v}). But a possible regularisation of the infinite integrals
would be the
introduction of a cut--off parameter $\Lambda$. This regularisation is
often used in effective non--renormalizable theories, where the cut--off
determines the scale up to which the effective description is valid.
The standard example is the Nambu and Jona--Lasinio (NJL) model.

But such a kind of additional parameter violates the scaling properties
of the effective theory, which we have chosen to simulate QCD,
because of the appearance of a new scale.
In order to keep the wanted scaling behaviour one has to give the
cut--off parameter also a conformal weight of one unit by some
mechanism, as for example done for the NJL model in
\cite{Kusaka92,Ripka92}.

We simplify the procedure here and cancel all infinite $T=0$
contributions by hand. This means that we normalize the theory at tree
level. We will determine the parameters $C$ and $\chi_p$ in (\ref{v}) such
that the EP at $T=0$ reproduces the bag constant
$\cal B$ and
the glueball mass $m_\chi$:
\beg\label{bag}
{\cal B} = V_{\rm eff}(\chi=0) - V_{\rm eff}(\chi=\chi_0) \: ,
\ben
\beg\label{masstn}
m^2_{\chi} = M^2_{\chi} (T=0) = V_{\rm eff}^{\dpri}(\chi=\chi_0) \: .
\ben
As we will see later the determination of the EP at
the origin $\bar{\chi} =  0$ is not possible because of infinities
in the effective coupling constant in the expansion of the
logarithm. Thus a renormalisation, i.e.~a determination of the
parameters via (\ref{bag}) is not possible.

By applying (\ref{bag}) and (\ref{masstn}), we fix the depth and the 
curvature at the minimum of the EP at $T=0$. We are dominantly interested 
in the thermal behaviour and therefore
in the thermal excitation around this minimum. But the characteristic
of this behaviour is already mainly fixed by these two parameters
(depth and curvature). Therefore we expect no drastic change in the
result of the thermal properties by changing the level of normalisation
in the theory.

For the value of the bag constant we choose as two limiting cases
${\cal B}^{1/4} = 140$ MeV and ${\cal B}^{1/4} = 240$ MeV which bound 
the common range of values appearing in the
literature. We want to emphasize that the larger value is more
realistic as newer results from hadron spectroscopy show
\cite{Hasenfratz81}, compared to the original work of the MIT--bag model
\cite{deGrand75} where the lower $\cal B$ was used.
Also, only the larger value is compatible with the
gluon condensate at $T=0$ as extracted from QCD sum rules
\cite{Shifman79}.

For the glueball mass we choose $m_{\chi} = 1.6$ GeV, but
we will also investigate other values. That value is motivated by
lattice calculations \cite{Bitar91}, and also the experimental search
favours candidates in the mass region of 1.5--1.8 GeV
\cite{databook}.

For the parameters of the chiral sector ($c$, $m_q$) we choose
standard values which reproduce the physical pion mass and, for the
case of SU(3)${}_{\rm b}$, also the $K$ and $\eta$ masses in a reasonable
way. We set $f_{\pi}= 93$ MeV.

We now apply (\ref{legen})--(\ref{veff2}) to the partition functions
(\ref{zchi})--(\ref{zsud}) and neglect as discussed the $T=0$ contributions 
in the integrals $I_n$. Up to first order in the expansion of
${\cal L}_{\rm int}^{\Omega}$ we are allowed to set
\cite{Iliopoulos75}
\beg
\bar{\chi} = \chicl \; .
\ben
For the case of a pure dilaton field theory we get for the GEP 
\beg\label{vchi}
   V^{\chi}_{\rm eff} (\ovchi) = V(\ovchi)  +
   I^T_1(\Omega^2,\,T)
   + \frac{1}{2} (V^{\dpri}(\ovchi) - \Omega^2) I^T_0(\Omega^2,\,T)
   + \; \frac{1}{8} V^{\fpri}(\ovchi) I^T_0(\Omega^2,\,T)^2 \; ,
\ben
and the variational equation for the determination of $\Omega$ reads
\beg\label{bchi}
   \frac{1}{2} (V^{\dpri}(\ovchi) - \Omega^2)
   + \; \frac{1}{4} V^{\fpri}(\ovchi) I^T_0(\Omega^2,\,T) = 0 \;.
\ben
The result is plotted in Fig.~2 for various temperatures and for the two
different bag constants. In that figure the singular region at 
$\ovchi=0$ has been truncated. We see from
eq.~(\ref{vchi}) that the four--point coupling of the $\chih$--field,
which is proportional to $V^{\fpri}(\ovchi) =
24\,C\,\left[\ln\left(\chi/\ovchi\right) + 11/6\right]$,
develops a singularity at
$\ovchi = 0$. The expansion of the tree level potential breaks down
at the origin. According to the interpretation of the model, the phase
$\ovchi=0$ would correspond to a vanishing gluon condensate, and
because of eq.~(\ref{qqcam}) also to a vanishing chiral condensate. This
means that all of the non--perturbative physics has gone and we are in the
pure perturbative region. An interpretation of this phase as the
deconfinement phase as done in \cite{Campbell90} is therefore natural.
But this would mean that the effective degrees of freedom of our
model, mesons and glueballs, have been dissolved. Therefore a breakdown
of the model at $\ovchi =0$ is only to be expected.

However, we can still use Fig.~2 to extract the temperature scale, where 
thermal excitations become important, even if we can't explicitly 
follow the phase transition to $\ovchi=0$. This temperature scale is 
apparently determined by the chosen bag constant. For 
${\cal B}^{1/4} = 140$ MeV we begin to see a strong shift of the
minimum at temperatures of about $T=250$ MeV, and above $T=300$ MeV the
minimum at $\ovchi \neq 0$ is lost, while for the large bag constant 
${\cal B}^{1/4} = 240$ MeV the GEP doesn't show any visible shift up to
temperatures of order $T \cong 350$ MeV, and only above $T \cong 450$ MeV 
the minimum at $\ovchi \neq 0$ disappears.

We now proceed to the GEP of the full Lagrangian (\ref{l}), where we
choose a SU($N$) flavour symmetric mass matrix ${\cal M}_q$. The result
for the GEP is
\begin{eqnarray}
   V^{{\rm SU(N)}}_{\rm eff} (\ovchi)& \! \! = \! \! &V^{\Phi}(\ovchi)  +
   I^T_1(\Omega^2,\,T)
   + \frac{1}{2} [(V^{\Phi})^{\dpri}(\ovchi) - \Omega^2] I^T_0(\Omega^2,\,T)
   + \; \frac{1}{8} V^{\fpri}(\ovchi) I^T_0(\Omega^2,\,T)^2
\nonumber\\
&& + (N^2-1) I^T_1(M^2_{\pi}(\ovchi),\,T)
   + \: \frac{(N^2-1) m_{\pi}^2}{4N\fpi} \left(\frac{\chi_0}{\ovchi}\right)
       I^T_0(M^2_{\pi}(\ovchi),\,T)^2 \nonumber \\
&& + \: \frac{(N^2-1) m_{\pi}^2}{2\ovchi\chi_0}  I^T_0(M^2_{\pi}(\ovchi),\,T)
       I^T_0(\Omega^2,\,T)  \: , \label{vsuz}
\end{eqnarray}
with the condition for $\Omega$
\beg\label{bsuz}
   \frac{1}{2} [(V^{\Phi})^{\dpri}(\ovchi) - \Omega^2] 
   + \; \frac{1}{4} V^{\fpri}(\ovchi) I^T_0(\Omega^2,\,T)
   + \frac{(N^2-1) m_{\pi}^2}{2\ovchi\chi_0}  I^T_0(M^2_{\pi}(\ovchi),\,T)
       I^T_0(\Omega^2,\,T) = 0 \;.
\ben
The result is plotted for $N=2$ in Fig.~3. Qualitatively the GEP behaves 
similary to the pure dilaton theory, Fig.~2. The dominant contribution to 
the shape of the GEP (\ref{vsuz}) as a function of $\ovchi$ comes from
the one loop term $I^T_1[\Omega^2(\ovchi),\,T]$ of the glueball
field. (This does not imply the same for the absolute values of the 
different terms in (\ref{vsuz}), because the overall normalisation 
of the curve in Fig.'s 2, 3 was chosen by hand to facilate comparison at 
different temperatures.) This means that the glueball dynamics itself
dominates the position of the minimum of the GEP. The Goldstone bosons and 
their coupling to the dilaton field play only a minor role.

The leading singularity at the origin, however, is influenced strongly by
the four--point coupling of the mesons and the coupling between glueballs 
and mesons which diverge as $1/\ovchi$ as $\ovchi \rar 0$. 
Therefore the two last terms in (\ref{vsuz}) dominate the singular 
behaviour and change the sign of the pole (now positive) relative to the 
pure glueball case. These explains the
steep rise of the EP curves near the origin at high temperatures.

As a final example we investigate the case of broken flavour 
SU(3)${}_{\rm b}$.
The result for the GEP is
\begin{eqnarray}
   V^{{\rm SU(3)_f}}_{\rm eff} (\ovchi)& \! \! = \! \! &V^{\Phi}(\ovchi)  +
   I^T_1(\Omega^2,\,T)
   + \frac{1}{2} [(V^{\Phi})^{\dpri}(\ovchi) - \Omega^2] I^T_0(\Omega^2,\,T)
   + \; \frac{1}{8} V^{\fpri}(\ovchi) I^T_0(\Omega^2,\,T)^2 \nonumber\\
&&   3 I^T_1(M^2_{\pi}(\ovchi),\,T)  + 4 I^T_1(M^2_{K}(\ovchi),\,T)
     + I^T_1(M^2_{\eta}(\ovchi),\,T) \nonumber \\
&&   + \frac{1}{24\fpi} \left(\frac{\chi_0}{\ovchi}\right)
     \Big[ 9m_{\pi}^2 I^T_0(M^2_{\pi}(\ovchi),\,T)^2
           - (m_{\eta}^2 + 4cm_s) I^T_0(M^2_{\eta}(\ovchi),\,T)^2 \nonumber\\
&&\hspace{2.7cm} - \: 6m_{\pi}^2 I^T_0(M^2_{\pi}(\ovchi),\,T)
     I^T_0(M^2_{\eta}(\ovchi),\,T) \nonumber\\
&&\hspace{2.7cm}+ \:16m_{K}^2 I^T_0(M^2_{K}(\ovchi),\,T)
     I^T_0(M^2_{\eta}(\ovchi),\,T)
      \Big] \nonumber\\
&& +  \frac{1}{2\ovchi\chi_0} I^T_0(\Omega^2,\,T)
      \Big[ 3m_{\pi}^2 I^T_0(M^2_{\pi}(\ovchi),\,T)
            +   4m_{K}^2 I^T_0(M^2_{K}(\ovchi),\,T) \nonumber\\
&&\hspace{3.3cm} + \: m_{\eta}^2 I^T_0(M^2_{\eta}(\ovchi),\,T)
        \Big] \; ,\label{vsud}
\end{eqnarray}
with
\begin{eqnarray}
&&  \frac{1}{2} [(V^{\Phi})^{\dpri}(\ovchi) - \Omega^2] 
   + \; \frac{1}{4} V^{\fpri}(\ovchi) I^T_0(\Omega^2,\,T)\nonumber\\
&& + \; \frac{1}{2\ovchi\chi_0}
      \Big[ 3m_{\pi}^2 I^T_0(M^2_{\pi}(\ovchi),\,T)
            +   4m_{K}^2 I^T_0(M^2_{K}(\ovchi),\,T)
            +  m_{\eta}^2 I^T_0(M^2_{\eta}(\ovchi),\,T) \Big] = 0 \;.
\label{bsud}
\end{eqnarray}
Here we only have a result for the large value of the bag constant, 
shown in Fig.~4. The reason is that for the potential $V^{\Phi}(\chi)$ 
(\ref{vphi}) the negative curvature due to the $g_{\chi}\chi^3$--term is 
so strong that by varying $C$ and $\chi_p$ the bag constant can never be
reduced below the value ${\cal B}^{1/4} = 220$ MeV. The general behaviour 
of the curves in Fig.~4 is similar to the former cases. The onset of the 
thermal shift of the minimum occurs at somewhat lower temperatures;
this can be seen better below, when we consider the gluon condensate.

\secspa \noindent
{3. CONDENSATES}\\
\underline{3.1.~The Gluon Condensate}\\
The minimum of the GEP determines the vacuum expectation value
$\langle \chi \rangle$. This minimum can only be determined numerically.
We then use the identification (\ref{i}) and set
\beg
\langle \frac{\beta(g)}{2g} G_{\mu \nu}^{\,a} G^{\,a \, \mu \nu} \rangle
= -C \langle \chi^4 \rangle  = -C\langle \chi \rangle^4 \; ,
\ben
where we use the approximation  $\langle O^n \rangle \approx
\langle O \rangle^n$, because the EP allows only the determination
of $\langle \chi \rangle$. The result is shown in Fig.~5 for different 
parameter sets.

The crucial point is that the gluon condensate is stable up
to temperatures of order 200 MeV for both bag constants. We already mentioned
that only the higher bag constant is realistic, where the decrease of the 
gluon condensate sets in even later. However, we must emphasize that our
results should be taken only on a very qualitative level above 
$T = 150$--200 MeV. The reason is that our starting point, 
the non--linear sigma model, represents only the lowest order of chiral
perturbation theory, and the neglecting of higher order gradient terms
limits its quantitative applicability to temperatures below the pion mass. 
Thus the relevant change in the gluon condensate happens only at temperatures
where the chiral sector of the model has already broken down.
However, the critical temperature for the melting of the
gluon condensate is dominated by the glueball dynamics, as can be seen
from a comparison in Fig.~5 of the pure glueball scenario with the model
including all eight pseudoscalar mesons. For this reason we believe
that our statements about the behaviour of the gluon condensate at
temperatures above the limit of validity of chiral perturbation
theory are at least qualitatively correct, unless the chiral phase 
transition implies also deconfinement for the glueballs
(for which our effective model does not provide any mechanism).

We see in Fig.~5 that the value for $T_c$ is dominated by the value of
the bag constant. A larger bag constant results in a higher $T_c$. 
At the high--temperature end of the drop in the gluon condensate the curves
seem to level off; thus the gluon condensate doesn't completely vanish 
at $T_c$, but approaches zero only in the limit $T\rightarrow \infty$. 
The surviving condensate just above the steep drop seems to be largest
for the case of SU(3)${}_{\rm b}$ with a large strange quark mass. This
ties in with the observation made in the introduction that the gluon 
condensate is strongly correlated to the scale anomaly. The anomaly, a 
quantum effect, is not expected to vanish at some fixed temperature, 
in contrast to spontaneously broken symmetries \cite{Dolan74,tanomaly}. 
What we apparently see here is the fading of the condensate relative to 
a new scale brought in by the temperature.

To be more specific, let us for simplicity look at the pure glueball
case. The scale of the anomaly is given by the gluon condensate
at $T=0$ or by the bag constant, which are related in our pure 
glueball model by
\beg
{\cal B} = - \; \frac{1}{4}
\langle \frac{\beta(g)}{2g} G_{\mu \nu}^{\,a} G^{\,a \, \mu \nu} 
\rangle_{T=0} \; .
\ben
We now read off the critical temperatures for the two investigated 
bag constants, obtaining $T_c=290$ MeV for ${\cal B}^{1/4} = 140$ MeV and
$T_c=470$ MeV for ${\cal B}^{1/4} = 240$ MeV. We thus have
\beg
T_c \approx 2 \times {\cal B}^{1/4} = \sqrt{2}\:
\langle \frac{\beta(g)}{2g} G_{\mu \nu}^{\,a} G^{\,a \, \mu \nu}
\rangle^{1/4}_{T=0} \; .
\ben
The result $T_c = {\cal O}(\langle G^2\rangle^{1/4}_{T=0})$ is no
surprise, since there is no other relevant scale. (The glueball mass is of 
minor influence, as will be shown below.) However, it confirms the 
interpretation, discussed above, that at $T_c$ the temperature becomes 
the dominating scale, leading to a fading of the gluon condensate. But no
complete vanishing is observed, meaning the anomaly stays on at high
temperatures. 
Therefore the non--perturbative physics incorporated
in the gluon condensate doesn't vanishes at $T_c$ but stays on to very
high temperatures. A continuation of non--perturbative physics beyond the
$T_c$ of the Wilson loop is
also seen in other sectors, e.g.~instanton effects \cite{Gross81} or
chromo--magnetic correlations \cite{magkorr}.

If we look at the influence of the chiral dynamics on the behaviour of
the gluon condensate, we see only minor effects. The general behaviour
is that as more mesons are incorporated as more $T_c$ is lowered, but the
corresponding shift is very small. A larger effect is seen when we include
flavour symmetry breaking in the SU(3) case by turning on the strange
quark mass and bringing the kaons and $\eta$ to their physical mass. 
This effect is at first surprising, because naively the heavier mesons 
shouldn't be excited as easily as in the case of exact SU(3) symmetry
with (nearly) massless quarks. The reason for the observed effect is 
the scaling behaviour of the meson masses (\ref{skalmass}), as we will 
now discuss.

We determine the gluon condensate via the minimum of the EP.
We vary $\ovchi$ and determine the energy density.
Important are, however, not the absolute
contributions, but the variation with $\ovchi$, because the global
normalisation of the EP is arbitrary. All meson masses scale with
$\sqrt{\ovchi/\chi_0}$. Therefore in the thermal weights we encounter 
expressions like
\beg
{\rm e}^{-m\sqrt{\ovchi/\chi_0}/T} \; ,
\ben
which vary more strongly with $\ovchi$ the larger the rest mass $m$ is.
Thus the location of the minimum is more sensitive to the heavier 
chiral mesons.

This, of course, raises the question of the influence of all the 
neglected other heavy hadrons. They surely are important for temperatures
of order 200 MeV and higher. Their influence would probably be a shift of 
$T_c$ to lower values, as indicated above. But the incorporation
of these other states into the model is problematic because for these 
non--Goldstone particles the scaling properties with $\ovchi$ are
not known. This limits the quantitative power of predictions of our 
model in the high temperature region.

We summarize by stating that the gluon condensate is very stable up
to temperatures above 200 MeV. This is in agreement with other results
for the temperature (in--)dependence of the gluon condensate within 
effective low--energy models \cite{Jaminon93,Gkonchi}. The only exception is
the work by Bernard et al.~\cite{Bernard89}, but this is probably due
to the very low bag constant taken in their approach.

We now vary the glueball mass at $T=0$ in order to study the influence
of this second empirical value on our results. The gluon condensate
as a function of $T$ is shown in Fig.~6 for three values of $m_{\chi}$. 
We see that a lower glueball mass drives $T_c$ to lower values, since
lighter glueballs are more easily excited at a given temperature, thus 
melting the condensate earlier. Quantitatively, however, the
effect is small: a variation of the glueball mass by 1 GeV (from 1 GeV to 
2 GeV) shifts $T_c$ only by about 50 MeV. Thus the dominant scale for 
$T_c$ is indeed the bag constant vis.~the gluon condensate at $T=0$.

\vspace*{2ex} \noindent
\underline{3.2.~The Quark Condensate}\\
We finally investigate the influence of the coupling between dilaton 
and chiral fields on the chiral condensate. In QCD the chiral
condensate is obtained by taking the derivative of the partition function
with respect to the current quark masses,
\beg
\langle q \bar{q} \rangle = \frac{\partial}{\partial m_q} Z^{\rm QCD}\; .
\ben
We replace the QCD generating functional by the effective
partition function (\ref{zchi})--(\ref{zsud}). We can use
eq.~(\ref{mesonmasses}) and rewrite the derivative with respect to
$m_q$ into a derivative with respect to $m_{\pi}$. This is equivalent to
using the Gell--Mann--Oakes--Renner relation \cite{Gellmann}
\beg\label{gor}
2\fpi^2m_{\pi}^2 = -(m_u + m_d)
\langle 0 | u\overline{u} + d\overline{d} | 0 \rangle + {\cal O}(m_q^2) \; ,
\ben
which connects the QCD parameters $m_q$ to the empirical parameters 
$f_{\pi}$, $m_{\pi}$ and $\langle q \bar{q} \rangle$ at zero temperature.
If we apply $\partial / \partial m_{\pi}$ to (\ref{zchi})--(\ref{zsud}),
we need to account for the implicit dependence of $\chicl$ on $m_{\pi}$ via
eq.~(\ref{cleom}), which results in the equation
\beg
(V^{\Phi})^{\prime}(\chicl) = 0 \; .
\ben
Using this condition we get
\beg\label{mincl}
\frac{\partial \chicl}{\partial m_{\pi}} =
-2 \left(\frac{\chicl}{\chi_0}\right)^2 \frac{m_{\pi} \fps}{M_{\chi}^2\chi_0}
\approx 6\cdot 10^{-3} \;\;\mbox{for} \;\chicl = \chi_0 \: .
\ben
Therefore we can neglect this dependence. Similarly the variational equation 
(\ref{bsuz}) causes a dependence of $\Omega$ on $m_{\pi}$. It is of the same 
order as (\ref{mincl}), and therefore we neglect it, too. We thus get for 
the chiral condensate
\newcommand{\spchi}{0.5ex}
\begin{eqnarray}
\frac{\langle q\overline{q}\rangle^{\rm SU(N)}(T)}{
  \langle q\overline{q}\rangle (T=0)} \!\!\!
&=& \!\!\!\Chicl^3 \Bigg[ 1 -
    \frac{N^2-1}{N\fps}\left(\frac{\chi_0}{\chicl}\right)^2 \;
                                               I^T_0(M_{\pi}^2(\chicl),\,T)
\nonumber\\[\spchi]
&& \!\!\!
  - \frac{N^2-1}{2N^2\fpi^4}\left(\frac{\chi_0}{\chicl}\right)^4 \;
                                             I^T_0(M_{\pi}^2(\chicl),\,T)^2
\nonumber\\[\spchi]
&& \!\!\!
    -\:  \frac{(N^2-1)\chi_0^2}{N\fps\chicl^4} \;I^T_0(M_{\pi}^2(\chicl),\,T)
                                                  I^T_0(\Omega^2,\,T)
    + \frac{3}{\chicl^2} \; I^T_0(\Omega^2(\chicl),\,T)
\nonumber\\[\spchi]
&& \!\!\!
    +\:  \frac{(N^2-1)m_{\pi}^2}{2N^2\fpi^4} \;
         \left(\frac{\chi_0}{\chicl}\right)^3 \;
                                      I^T_0(M_{\pi}^2(\chicl),\,T)
                                  I^T_{-1}(M_{\pi}^2(\chicl),\,T)
\nonumber\\[\spchi]
&& \!\!\!
    +\: \frac{(N^2-1)m_{\pi}^2\chi_0}{2N\fps\chicl^3} \; I^T_0(\Omega^2,\,T)
                                  I^T_{-1}(M_{\pi}^2(\chicl),\,T)
     \Bigg] \; .
\label{qqsuz}
\end{eqnarray}

If we set $(\chicl / \chi_0) = 1$ and cancel all terms containing the
dilaton mass parameter $\Omega$, we recover the well known result of
chiral perturbation theory \cite{leutwyler}. The corrections
due to the dilaton field are of two kinds. First we have the
universal scaling by $(\chicl / \chi_0)^3$ as already shown in
eq.(\ref{qqcam}). Second also the $\langle {\rm Tr}(U + U^{\dag}) \rangle$ 
factor receives corrections because of the coupling between the dilaton 
and the chiral field. First there is a scaling of the terms coming from pure
meson loops because of the scaling of the meson propagators
with negative powers of $(\chicl / \chi_0)$ (\ref{prophi}). Second there
are contributions from glueball loops. These last terms are small as
has already been shown in \cite{Dotterweich93}.

The quantitativly largest modification comes from the universal scaling,
which goes with the third power of $\chicl$. The result is plotted in 
Fig.~7 for a flavour SU(2), where we set $\chicl = \bar{\chi}$ and
use the temperature dependence of $\bar{\chi}$ and $\Omega$
as extracted from the GEP. The solid line corresponds to the result of
chiral perturbation theory \cite{leutwyler}. The modifications are really
small. For the realistic larger bag constant there is nearly no visible
shift, because $(\ovchi/\chi_0)\approx 1$ in the whole temperature region
where the chiral condensate is non--zero. Only the smaller bag constant 
leads to a downward shift of order 20 MeV of the chiral $T_c$.
In general we conclude that the scaling properties of QCD have only a
very weak influence on the chiral dynamics.

\secspa \noindent
{4. CONCLUSIONS}\\
We have started from an effective Lagrangian which implements the chiral
symmetry and scaling aspects of QCD. This allows the investigation of two
interesting questions, the thermal behaviour of the gluon condensate
and the influence of the scaling properties on the well investigated
chiral restoration phenomenon.

We find that the gluon condensate is very stable up to temperatures of 
200 MeV, where the chiral sector of the theory reaches its limit of
validity. As a result the chiral dynamics is hardly changed at all up to 
the chiral phase transition. Thermal variations of the gluon condensate 
and of the chiral condensate occur at two quite different
temperature scales. A conclusion from our work is that the gluon 
condensate does not drive the disappearance of the chiral condensate at 
high temperature as suggested as one possible scenario in \cite{Campbell90}.

In our model the melting of the gluon condensate is dictated by the
glueball dynamics itself. The influence of the chiral mesons is minor,
but other hadron states neglected in our approach may further change 
the temperature scale of the gluon condensate. This thermal scale is 
fixed by the value of the condensate at zero temperature or, equivalently,
the bag constant. In a pure glueball theory we find
$T_c^{\langle GG\rangle} \approx \sqrt{2} \langle GG\rangle^{1/4}_{T=0}$.

It was suggested in \cite{Campbell90} that the expectation value 
$\langle \chi\rangle$ of the dilaton field, i.e.~the gluon condensate,
could be used as an order parameter for gluon deconfinement. Our
results contradict this interpretation. We see a relatively stable gluon 
condensate, well beyond the point of evaporation of the quark condensate
which, according to \cite{Campbell90}, should be interpreted as quark
deconfinement. Since we see no reason why one kind of colored particle 
should be deconfined earlier than another one, we suggest that the color
deconfinement phase transition occurs before the gluon condensate vanishes.
This implies that nonperturbative phenomena persist well into the deconfined
phase.

This is also seen in lattice QCD results, where one is able to extract
via some ex\-tra\-po\-la\-tion the finite temperature behaviour of the gluon
condensate \cite{lattice}. There one sees at the critical temperature
$T_c^{\rm WL}$ of the Wilson loop no evidence for a rapid decrease of 
the gluon condensate. The authors of \cite{lattice} suggest that at 
$T_c^{\rm WL}$ half of the zero temperature condensate survives. All of 
these analysis suggest that the gluon condensate is not suited as an 
order parameter for deconfinement.

This conclusion is also supported by the fact that the gluon condensate
is strongly connected to the QCD scale anomaly. Quantum--field theoretical
anomalies are not expected to vanish at some finite temperature or density. 
Our results for the gluon condensate seem to support
this expectation, since in our calculation the condensate, after
a steep decrease at some $T_c$, levels off at a finite value, suggesting
complete disappearance only in the limit $T\rightarrow \infty$.

\secspa \noindent
{\bf Acknowledgement:}
We would like to thank K.~Kusaka, G.~Ripka, K.~Kajantie and W.~Weise 
for important comments and useful discussions. We acknowledge the support
of this work by BMFT (contract 06 OR 734), DFG (grant He 1283/3-2)
and GSI (contract OR Hei T). J.S.~greatfully acknowledges support by 
a fellowship of the Freistaat Bayern. 

\secspa \noindent
{\bf Appendix}\\
Here we present a list of integrals used in the text. We use the notation 
of \cite{Gauss}. The zero temperature integrals are defined by
\beg
I_N^{T=0} (M^2) =
\int\frac{\d^3 p}{2E(\vec{p})(2\pi)^3} [E(\vec{p})]^{2N} \; .
\ben
These and their finite temperature analogues satisfy the following
recursion relation:
\beg
\frac{\d I_N(\Omega^2)}{\d \Omega^2} = \frac{2N-1}{2} \;I_{N-1}(\Omega^2) \; .
\ben
In this paper we have used the following integrals:
\newcommand{\asp}{1.5ex}
\begin{eqnarray}
I_{-1}(M^2,\,T)  &=&I_{-1}^{T=0}(M^2) + I_{-1}^T(M^2,\,T)\\[\asp]
I_{-1}^{T=0}(M^2)&=&\int\frac{\d^3\vec{p}}{(2\pi)^3}\frac{1}{2E^3(\vec{p})}
\\[\asp]
I_{-1}^T(M^2,\,T)&=&\int\frac{\d^3\vec{p}}{(2\pi)^3} \frac{1}{E^2(\vec{p})
                    \:(e^{\beta E(\vec{p})}-1)}
                    \left[\frac{1}{E(\vec{p})} +
                    \frac{\beta e^{\beta E(\vec{p})}}{
                     e^{\beta E(\vec{p})}-1}\right] \nonumber \\
                 &=&\frac{1}{4\pi^2}\int\limits_{\beta M}^{\infty} \d x
                    \frac{\sqrt{x^2-\beta^2 M^2}}{x^2(e^x - 1)^2}
                    \left[(1+x)e^x -1\right] \\[\asp]
I_0(M^2,\,T)     &=&I_0^{T=0}(M^2) + I_0^T(M^2,\,T)\\[\asp]
I_0^{T=0}(M^2)   &=&\int\frac{\d^3\vec{p}}{(2\pi)^3} \frac{1}{2E(\vec{p})}
\\[\asp]
I_0^T(M^2,\,T)   &=&\int\frac{\d^3\vec{p}}{(2\pi)^3} \frac{1}{E(\vec{p})}
                    \:\frac{1}{e^{\beta E(\vec{p})}-1}
                  = \frac{T^2}{2\pi^2}\int\limits_{\beta M}^{\infty} \d x
                    \frac{\sqrt{x^2-\beta^2 M^2}}{e^x - 1} \\[\asp]
I_0^T(M^2=0,\,T) &=& \frac{T^2}{12} \\[\asp]
I_1(M^2,\,T)     &=&I_1^{T=0}(M^2) + I_1^T(M^2,\,T)\\[\asp]
I_1^{T=0}(M^2)   &=&\int\frac{\d^3\vec{p}}{(2\pi)^3} \frac{E(\vec{p})}{2}
\\[\asp]
I_1^T(M^2,\,T)   &=&\frac{1}{\beta}\int\frac{\d^3\vec{p}}{(2\pi)^3}
                   \:\ln\left(1-e^{-\beta E(\vec{p})}\right)\nonumber \\
                 &=& \frac{T^4}{2\pi^2}\int\limits_{\beta M}^{\infty} \d x
                  \, x \, \sqrt{x^2-\beta^2 M^2} \ln\left(1-e^x\right)\\[\asp]
I_1^T(M^2=0,\,T) &=& - \frac{\pi^2\,T^4}{90}
\end{eqnarray}

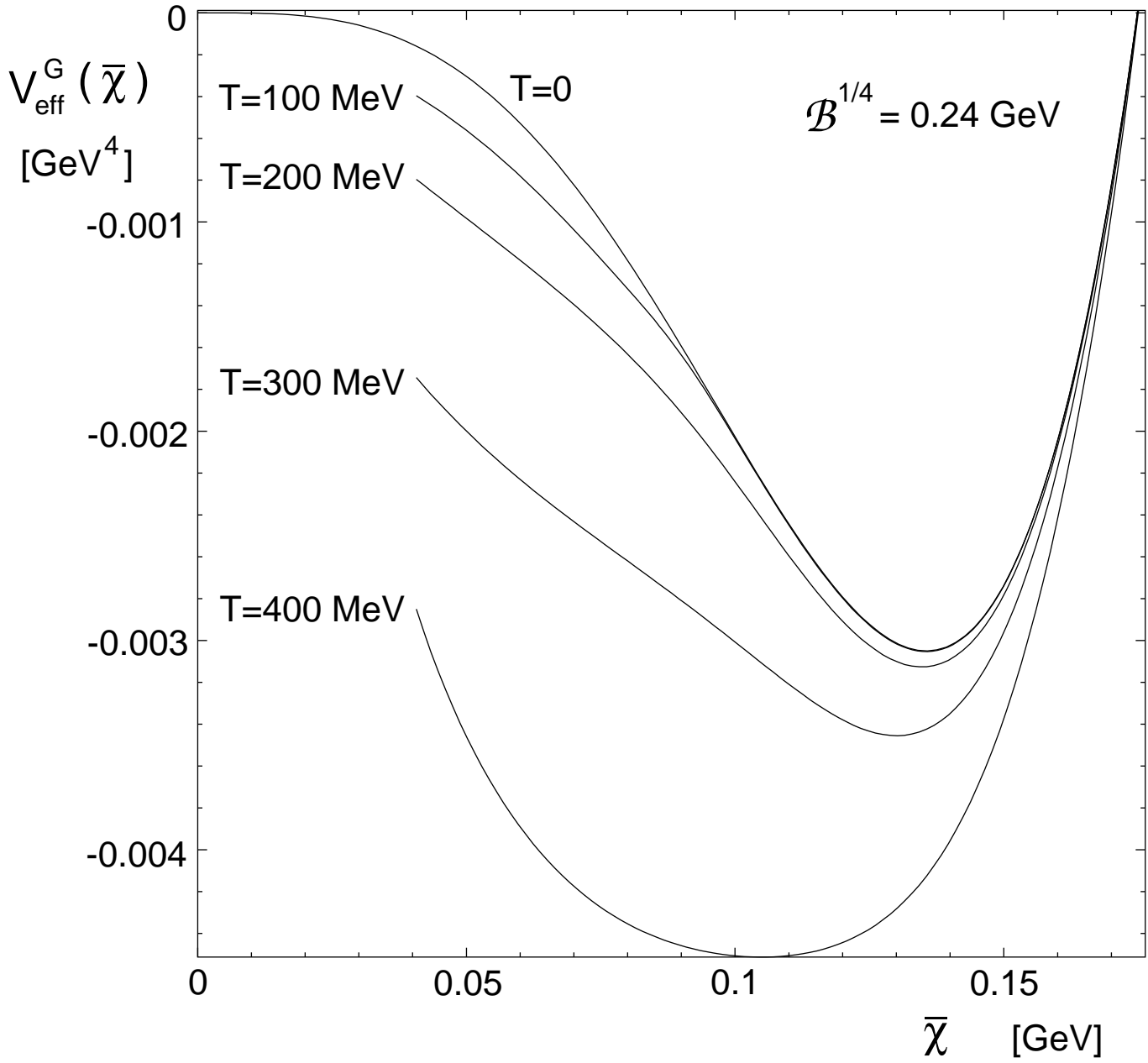
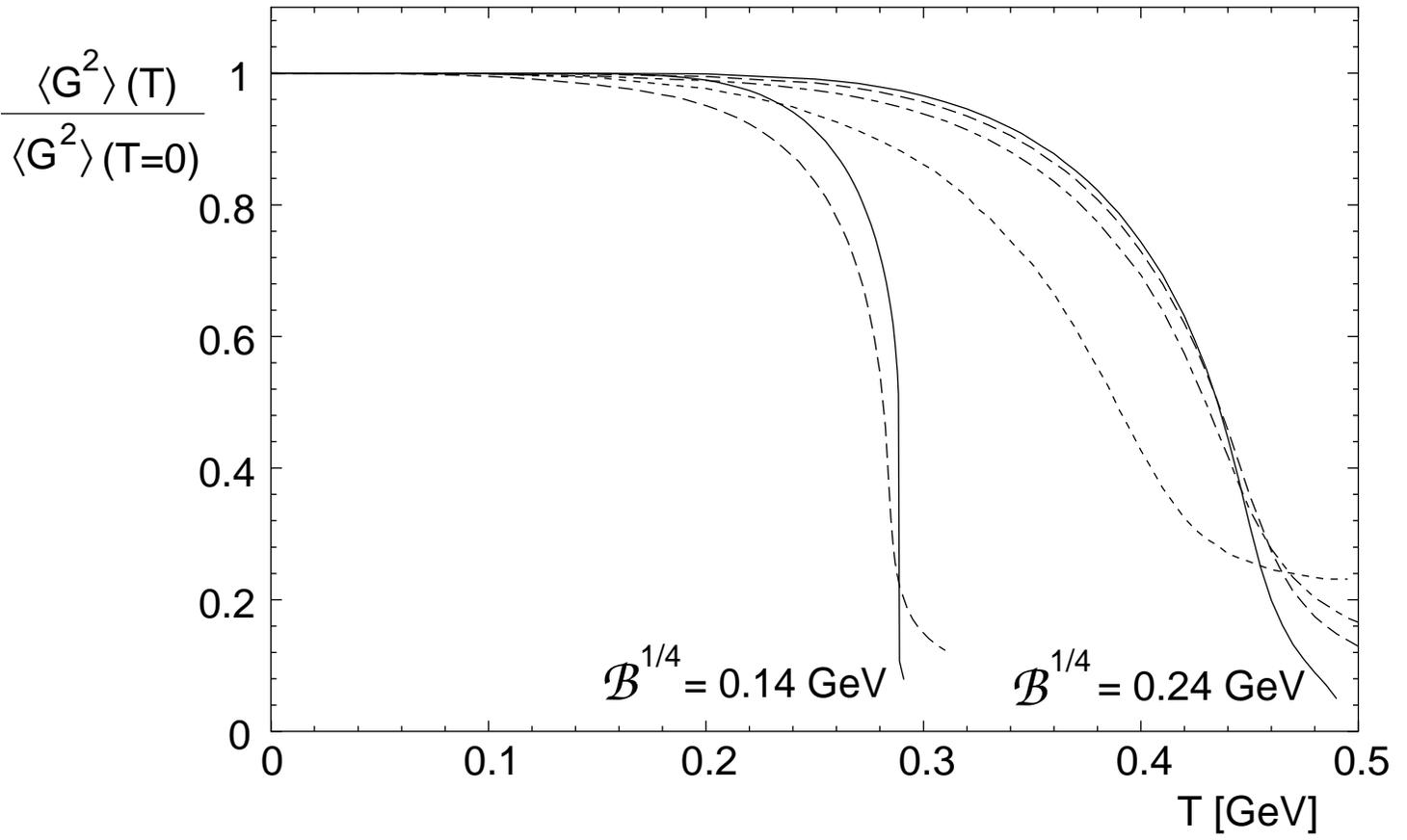
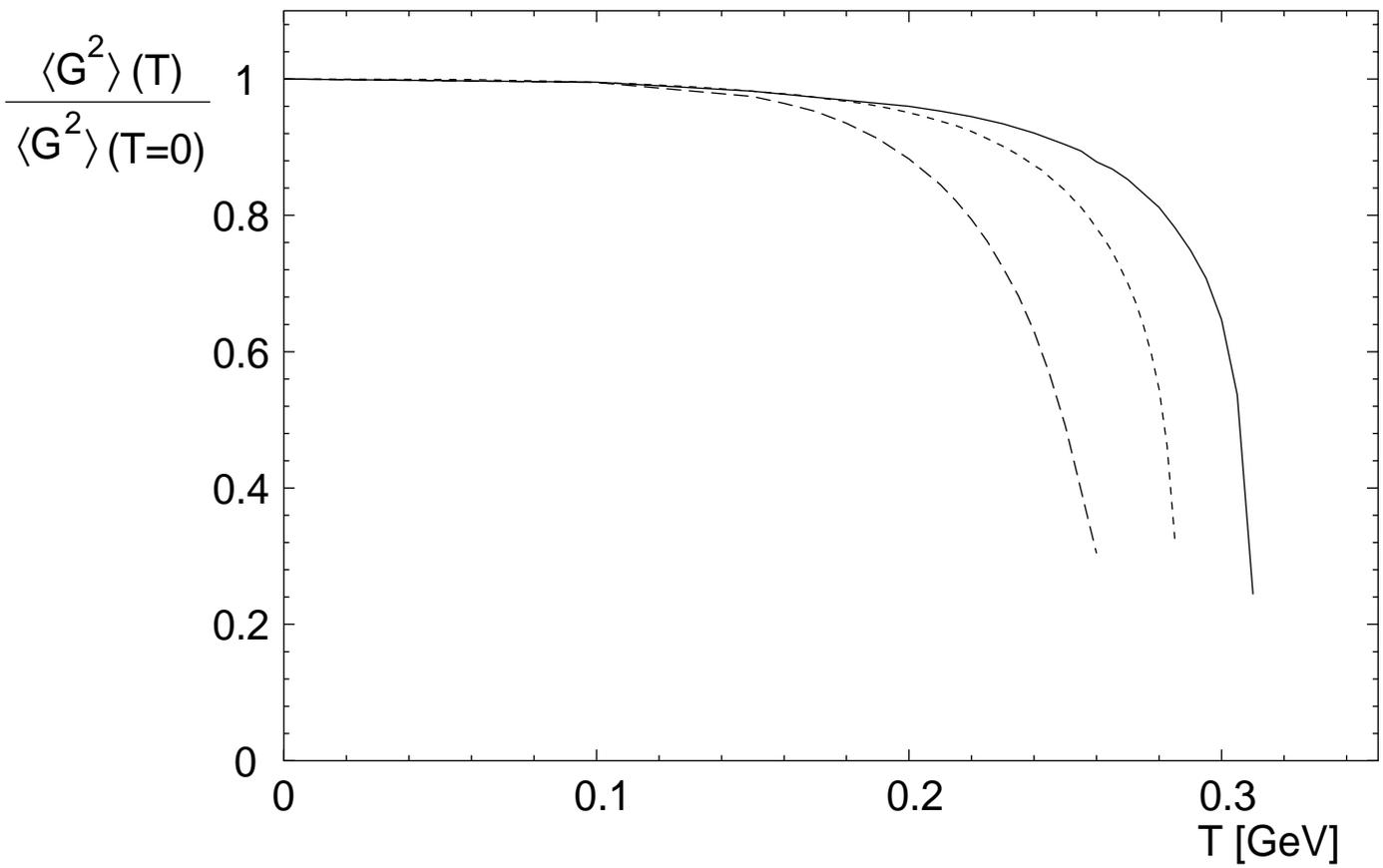
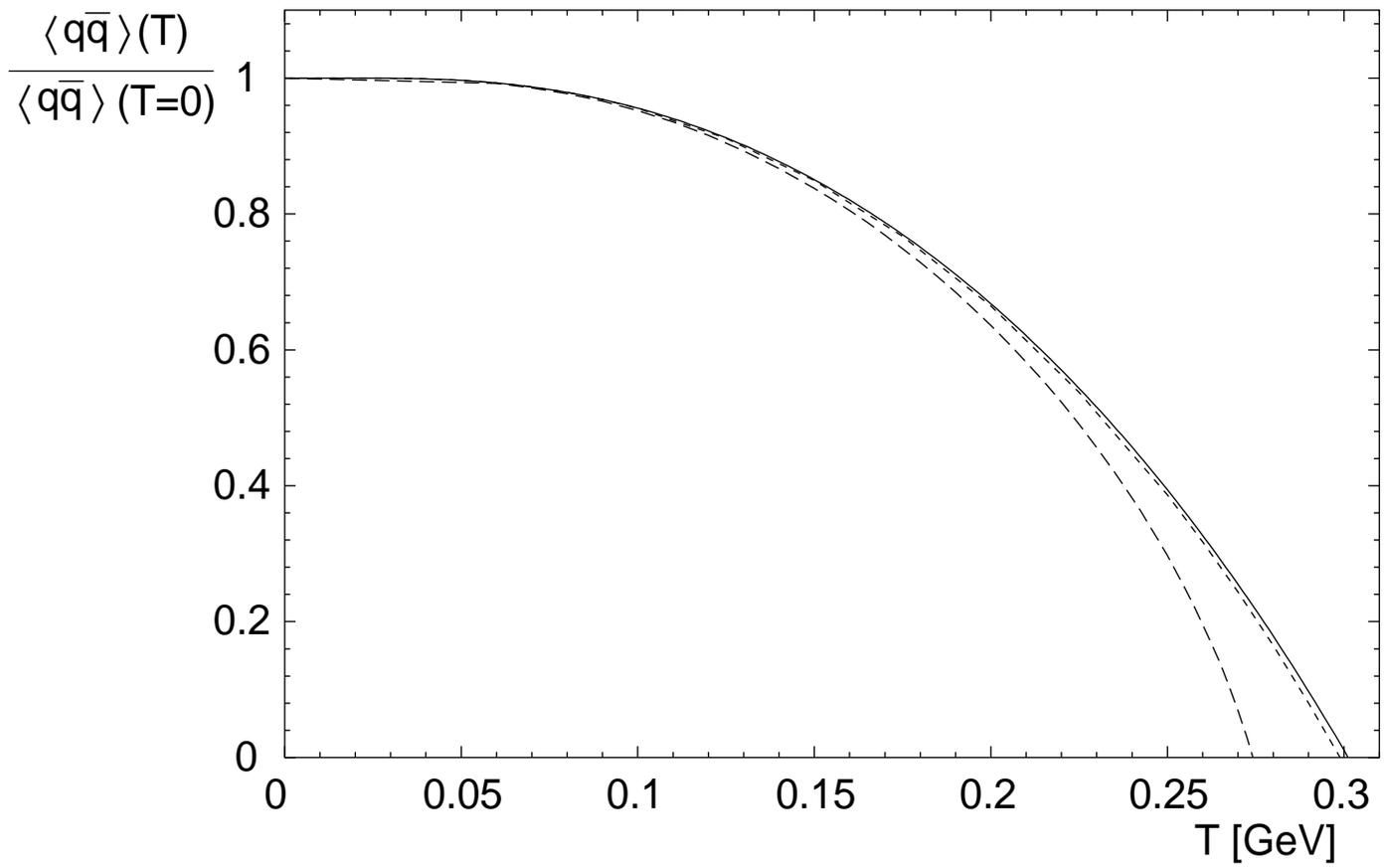
\end{document}